# Universal Statistics of the Scattering Coefficient of Chaotic Microwave Cavities

Sameer Hemmady[1,2], Xing Zheng[2], Thomas M. Antonsen Jr.[1,2], Edward Ott[1,2] and Steven M. Anlage[1,3].
*Department of Physics, University of Maryland, College Park, MD-20742.*

*(Dated: March 6$^{th}$, 2005)*

**Abstract:**

We consider the statistics of the scattering coefficient $S$ of a chaotic microwave cavity coupled to a single port. We remove the non-universal effects of the coupling from the experimental $S$ data using the radiation impedance obtained directly from the experiments. We thus obtain the normalized, complex scattering coefficient whose Probability Density Function (PDF) is predicted to be universal in that it depends only on the loss (quality factor) of the cavity. We compare experimental PDFs of the normalized scattering coefficients with those obtained from Random Matrix Theory (RMT), and find excellent agreement. The results apply to scattering measurements on any wave chaotic system.

**PACS# : 05.45.Mt, 03.65.Nk,11.55.-M,03.50.De,84.40.-X,84.40.Az**

## *I. Introduction:*

The scattering of short wavelength waves by chaotic systems has motivated intense research activity, both theoretically and experimentally [1–3]. Some examples of wave chaotic systems include quantum dots [4], atomic nuclei [5], microwave cavities [2, 6], and acoustic resonators [7]. Since these systems all have underlying chaotic ray dynamics, the wave pattern within the enclosure, as well as the response to external inputs, can be very sensitive to small changes in frequency and to small changes in the configuration. This motivates a statistical approach to the wave scattering problem.

The universal distribution for chaotic scattering matrices can be described by Dyson's circular ensemble [9, 10]. However, the circular ensemble cannot typically be directly compared with experimental data because it applies only in the case of 'ideal coupling' (which we define subsequently), while in experiments there are non-ideal, system-specific effects due to the particular means of coupling between the scattering system (e.g, a microwave cavity) and the outside world. This non-universality of the raw experimental scattering data has long been appreciated and addressed in theoretical work [11–13]. Of particular note is the work of Mello, Peveyra and Seligman (MPS) which introduces the distribution known as the Poisson kernel, where a scattering matrix $<S>$ is used to parameterize the non-ideal coupling. To apply this theory to an experiment it is typically necessary to specify a procedure for determining a measured estimate of $<S>$.

The first microwave experiments investigating scattering statistics by comparison with the theory of Refs. [11-13] were those of Ref. [15]. In analyzing their experimental data, the authors of Ref.15 considered the power reflection coefficient $|S|^2$ and used values of this quantity averaged over a number of configurations and over a suitable frequency range $(f_0 - \Delta f/2)$ to $(f_0 + \Delta f/2)$ to compare with theory. For a given number of averaging configurations, the chosen frequency range must be large enough that good statistics are obtained, but at the same time be small enough that the variation of the coupling with frequency is not significant. Subsequent work by the authors of Ref.[15,16], focused on the complex scattering amplitude. Again averages of measured values were made over a range of frequencies and a number of different configurations. We denote this experimentally obtained average as $\bar{S}$. If the frequency range is small enough and a sufficient number of independent measurements are obtained $\bar{S}$ will yield a good approximation to the desired $<S>$.

The quantity $<S>$ in the theory describes direct (or prompt) processes [12,14] that depend only on the local geometry of the coupling ports, as opposed to complicated chaotic processes resulting from multiple reflections far removed from the coupling port. On the other hand, in obtaining $\bar{S}$, as in Ref. [16], averaging of the scattering data of the full chaotic system is employed. Thus the data used to obtain $\bar{S}$ is the same measured data whose statistics are being studied. Also note that $<S>$ is presumed to characterize the coupling, which is independent of the chaos of the system and is thus, in principle, a non-statistical quantity. In this paper we shall pursue another approach [17,18]. Specifically we seek to characterize the coupling in a manner that is both independent of the chaotic system and obtainable in a non-statistical manner (i.e. without employing averages). As we explain in more detail subsequently, this latter point is of practical importance because of the inherent inaccuracy and sample size issues introduced by averaging a fluctuating quantity.

[1] Also with the Department of Electrical and Computer Engineering.
[2] Also with the Institute for Research in Electronics and Applied Physics.
[3] Also with the Center for Superconductivity Research.

As discussed in Ref. 17 and 18, a direct means of investigating the statistics of typical chaotic scattering systems can be based on determination of the radiation impedance of the port $Z_{rad}$ or equivalently $S_{rad}$, the complex radiation scattering coefficient. These will be discussed in more detail subsequently. The radiation scattering coefficient (respectively, radiation impedance) is the scattering coefficient (impedance) that would be observed if the distant boundaries of the cavity were made perfectly absorbing or moved to infinity. Therefore, it describes prompt processes at the port and can be shown to be equal to $<S>$. The perfect coupling case corresponds to $S_{rad} = 0$, in which all incident wave energy enters the cavity.

The radiation scattering coefficient $S_{rad}$ can be directly measured in microwave experiments without resorting to averaging over a range of frequencies. Note, that $S_{rad}$ depends on the microwave frequency and thus, for the purposes of taking into account coupling, $S_{rad}$ can be a more useful and robust means of extracting universal properties from the data than $\bar{S}$, which depends on the frequency range $\Delta f$ over which the averaging is done.

The impedance is another fundamental quantity characterizing coupling to a cavity is the impedance $Z$. It is related to $S$ through the bilinear transformation,

$$S = (Z + Z_o)^{-1}(Z - Z_o), \quad (1)$$

where $Z_o$ is the characteristic impedance of the transmission line or waveguide feeding the port. The impedance linearly relates the voltage and the current phasors at the port and is determined solely by properties of the cavity and its port. In what follows we only discuss the one-port case, hence $Z$ and $S$ are scalars (rather than matrices) throughout this paper. A more general discussion involving multiple ports can be found in [19]. Inverting the transformation Eq.(1), we can relate the radiation impedance $Z_{rad} = R_{rad} + iX_{rad}$ to the radiation scattering coefficient $S_{rad}$ as,

$$Z_{rad} = Z_o \frac{(1 + S_{rad})}{(1 - S_{rad})}. \quad (2)$$

In Ref.[18-20] it was shown that the cavity impedance $Z$ can be expressed in terms of a scaled impedance $z$ and the radiation impedance $Z_{rad}$ as,

$$Z = iX_{rad} + R_{rad}z, \quad (3)$$

where $z$ is a complex random variable satisfying $<z> = 1$. The random variable $z$ has universal properties and describes the impedance fluctuations of a perfectly coupled cavity. The real part of $z$ is well known in solid state physics as the local density of states (LDOS) and its statistics have been studied [21,22]. The imaginary part of $z$ determines fluctuations in the cavity reactance. Discussion of the probability distribution of the normalized impedance $z$ for microwave experiments has been presented in a previous paper [18].

The purpose of this paper is to study the universal statistical properties of the cavity scattering coefficient $s$ defined in terms of the normalized impedance $z$,

$$s = (z-1)/(z+1) = |s|e^{i\phi_s}, \quad (4)$$

which can be compared directly to theoretical predictions based on ideal coupling. Combining Eqs. (2), (3) and (4), we obtain a formula relating $s$ to the cavity scattering coefficient $S$ and the cavity radiation scattering coefficient $S_{rad}$,

$$s = \frac{(1 + S_{rad}^*)}{(1 + S_{rad})} \frac{(S - S_{rad})}{(1 - SS_{rad}^*)}. \quad (5)$$

The inverse of Eq.(5) giving the actual scattering amplitude $S$ in terms of the normalized scattering amplitude $s$ is a statement of the Poisson Kernel [11,12] for a single port cavity with internal loss. We note that the first factor in Eq.(5) is simply a phase shift which depends only on the coupling geometry. Thus, the magnitude of $s$ satisfies,

$$|s| = \left| \frac{S - S_{rad}}{1 - SS_{rad}^*} \right| \quad (6)$$

According to the statistical theory, the only parameter on which the statistics of $z$ and $s$ depend is the loss due to the cavity wall absorption, which can be controlled and quantified in our experiment. We will experimentally verify the key theoretical predictions based on the circular ensemble hypothesis, such as the statistical independence of the magnitude and phase of the normalized scattering coefficient $s$, and the uniform probability distribution for the phase. Different degrees of loss and different coupling structures will be examined, and approximations to the probability density functions of $s$ and $z$ will be compared with the theoretical predictions from Random Matrix Theory (RMT).

An expression similar to Eq.(6) was used by Kuhl *et al.*[16] to generate distributions of the scattering amplitudes $S$ based on theoretical predictions for the normalized scattering amplitude $s$. Thus, it was assumed that the normalized scattering amplitude had the predicted properties of independence of magnitude and phase, and uniform distribution of phase. We will experimentally test these assumptions by using Eq.(5) directly to determine the properties of $s$.

Our paper is organized as follows. Section II presents our experimental setup and data collection. Section III carries out the normalization process to recover universal scattering characteristics and presents experimental histogram approximations to the probability density functions of the magnitude and phase of the normalized scattering coefficient $s$ for different coupling structures and losses. Section IV explores a predicted relationship between the average cavity power reflection coefficient ($\langle |S|^2 \rangle$) and the magnitude of the radiation scattering coefficient. Section V discusses the advantages of employing the radiation scattering coefficient in uncovering universal properties, or of predicting raw scattering data. Section VI concludes the paper and gives the summary.

## *II. Experimental Setup:*

Microwave cavities with irregular shapes (having chaotic ray dynamics) have proven to be very fruitful for the study of wave chaos, where not only the magnitude, but also the phase of scattering coefficients, can be directly measured from experiments. Our experimental setup consists of an air-filled quarter bow-tie chaotic cavity (Fig.1(a)) which acts as a two dimensional resonator below about 19 GHz [23]. Ray trajectories in a closed billiard of this shape are known to be chaotic. This cavity has previously been used for the successful study of the eigenvalue spacing statistics [6] and eigenfunction statistics [24,25] for a wave chaotic system. In order to investigate a scattering problem, we excite the cavity by means of a single coaxial probe whose exposed inner conductor, with a diameter (2a) extends from the top plate and makes electrical contact with the bottom plate of the cavity (Fig.1(b)). In this paper we study the properties of the cavity over a frequency range of 6 - 12 GHz, where the spacing between two adjacent resonances is on the order of 25 – 30 MHz.

As in the numerical experiments in Refs.[18,19], our experiment involves a two-step procedure. The first step is to collect an ensemble of cavity scattering coefficients $S$ over the frequency range of interest. Ensemble averaging is realized by using two rectangular metallic perturbations with dimensions 26.7 x 40.6 x 7.87 mm³ (~1% of the cavity volume), which are systematically scanned and rotated throughout the volume of the cavity (Fig.1(a)). Each configuration of the perturbers within the cavity volume results in a different value for the measured value of $S$. This is equivalent to measurements on cavities having the same volume, loss and coupling geometry for the port, but with different shapes. The perturbers are kept far enough from the antenna so as not to alter its near-field characteristics. For each configuration, the scattering coefficient $S$ is measured in 8000 equally spaced steps over a frequency range of 6 to 12 GHz. In total, one hundred different configurations are measured, resulting in an ensemble of 800,000 $S$ values. We refer to this step as the "cavity case".

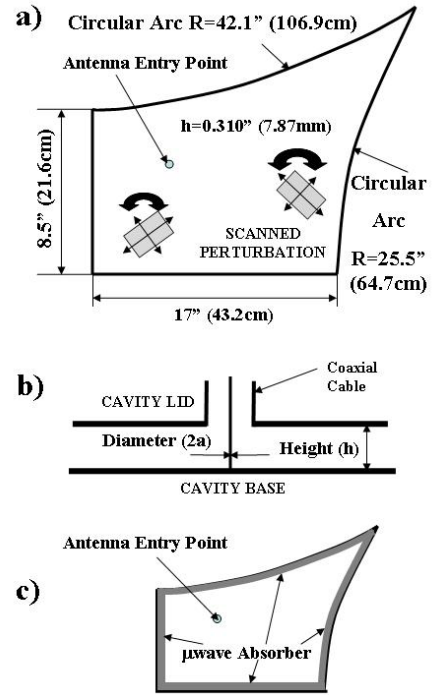

**Fig.1:** (a) The physical dimensions of the quarter bow-tie chaotic microwave resonator are shown along with the position of the single coupling port. Two metallic perturbations are systematically scanned and rotated throughout the entire volume of the cavity to generate the cavity ensemble. (b) The details of the coupling port (antenna) and cavity height $h$ are shown in cross section. (c) The implementation of the radiation case is shown, in which commercial microwave absorber is used to line the inner walls of the cavity to minimize reflections.

The second step, referred to as the "radiation case", involves obtaining the scattering coefficient for the excitation port when waves enter the cavity but do not return to the port. In the experiment, this condition is realized by removing the perturbers and lining the side walls of the cavity with commercial microwave absorber (ARC Tech DD10017D) which provides about 25dB of reflection loss between 6 and 12 GHz (Fig.1.(c)). Note that the side walls of the cavity are outside the near field zone of the antenna. Using the same frequency stepping of 8000 equally spaced points over 6 to 12 GHz, we measure the radiation scattering coefficient $S_{rad}$ for the cavity. Such an approach approximates the situation where the side walls are moved to infinity; therefore $S_{rad}$ does not depend on the chaotic ray trajectories of the cavity, and thus gives a characterization of the coupling independent of the chaotic system. Because the coupling properties of the antenna depend on the wavelength and thus vary over frequency, $S_{rad}$ is usually frequency dependent.

Having measured the cavity $S$ and $S_{rad}$, we then transform these quantities into the corresponding cavity and radiation impedances ($Z$ and $Z_{rad}$) respectively using Eqs. (1) and (2). The normalized impedance $z$ is obtained by Eq. (3). In order to obtain $z$, every value of the determined cavity impedance $Z$ is normalized by the corresponding value of $Z_{rad}$ at the same frequency. The transformation in Eq.(4) (or equivalently, Eq.(5)) yields the normalized scattering coefficient $s=|s|\exp(i\phi_s)$, which is the key quantity of interest in this paper. Since the artifacts of non-ideal coupling are supposed to have been "filtered out" through this normalization process, the statistics of the ensemble of $s$ values should depend only on the net cavity loss.

In order to test the validity of the theory for systems with varying loss, we create different "cavity cases" with different degrees of loss. Loss is controlled and parameterized by placing 15.2 cm-long strips of microwave absorber along the inner walls of the cavity. These strips cover the side walls from the bottom to top lids of the cavity. We thus generate three different loss scenarios (Loss Case 0, Loss Case 1, Loss Case 3) (shown schematically in the insets to Fig 4(b)). The numbers 0, 1, 3 correspond to the number of 15.2 cm -long strips placed along the inner cavity walls. The total perimeter of the cavity is 147.3 cm.

The theory predicts that as long as the loss is the same, the normalized $z$ or $s$ will have the same statistical behavior. This prediction will be tested in our experiments with two different coupling geometries corresponding to coaxial cables with two different inner diameters (2a=1.27mm and 2a=0.625mm, schematically shown in Fig.1(b)).

### III:  Experimental Results and Data Analysis:

In this section, we present our experimental findings for the statistical properties of the normalized scattering coefficient $s$, for different coupling geometries and degrees of loss. This section is divided into three parts. In the first part, we give an example for the PDF of $s$ at a specific degree of quantified loss and a certain coupling geometry. In the second part, we fix the degree of quantified loss, but vary the coupling by using coaxial cable antennas having inner conductors of different diameters (2a=1.67mm and 2a=0.635mm). The PDF histograms for the magnitude and phase of $s$ in these two cases will be compared. Finally, in the third part, we test the trend of the PDF of $|s|^2$ for a given coupling geometry and for three different degrees of quantified loss. Good agreement with random matrix theory is found in all cases.

### III (a):Statistical Independence of $|s|$ and $\phi_s$:

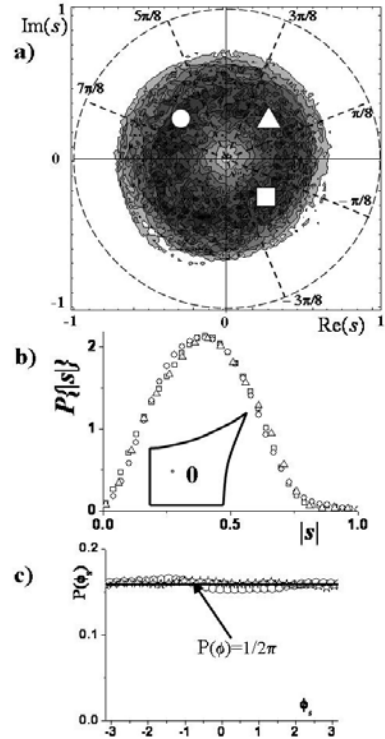

**Fig.2:** (a) Polar contour density plot for the real and imaginary components of the normalized cavity $s$ ($s=|s|\exp(i\phi_s)$) for Loss case 0 in the frequency range of 6 to 9.6 GHz. The angular slices with the symbols (triangles, circles, squares) indicate the regions where the PDF of $|s|$ is calculated and shown in (b). Observe that the PDF of the three regions are essentially identical. (c) The PDF of the phase $\phi_s$ of the normalized scattering coefficient $s$ for two annuli defined by $0\leq|s|\leq 0.3$ (stars) and $0.3<|s|\leq 0.6$ (hexagons). Observe that these phase PDFs are nearly uniform in distribution. The uniform distribution is shown by the solid line ($P(\phi)=1/2\pi$). This is consistent with the prediction that the $|s|$ is statistically independent of the phase $\phi_s$, of $s$.

The first example we give is based on loss case 0 (i.e., no absorbing strips in the cavity) and coupling through a coaxial cable with inner diameter 2a=1.27mm. Having obtained the normalized impedance $z$, we transform $z$ into the normalized scattering coefficient $s$ using Eq. (4). Since the walls of the cavity are not perfect conductors, the normalized scattering coefficient *s* is a complex scalar with modulus less than 1. (In Loss Case 0, most of the loss occurs in the top and bottom cavity plates since they have much larger area than the side walls.) Based on Dyson's circular ensemble, one of the most important properties of *s* is the statistical independence of the scattering phase $\phi_s$ and the magnitude $|s|$. Fig.2(a) shows a contour density plot of *s* in the frequency range of 6 to 9.6 GHz for Loss Case 0. The grayscale level at a given point in Fig. 2(a) indicates the number of points (for {Re(*s*), Im(*s*)}) that fall within a local rectangular region of size 0.02 X 0.02. We next arbitrarily take angular slices of this distribution that subtend an angle of π/4 radians at the center, and compute the histogram approximations to the PDF of $|s|$ using the points within those slices. The corresponding PDFs of $|s|$ for the three slices are shown in Fig.2(b). We observe that these PDFs are essentially identical, independent of the angular slice. Fig.2(c) shows PDFs of $\phi_s$ computed for all the points that lie within two annuli defined by $0 \leq |s| \leq 0.3$ (stars) and $0.3 < |s| \leq 0.6$ (hexagons). These plots support the hypothesis that the magnitude of *s* is statistically independent of the phase $\phi_s$ of *s* and that $\phi_s$ is uniformly distributed in 0 to 2π. To our knowledge, this represents the first experimental test of Dyson's circular ensemble hypothesis for wave chaotic scattering.

### III(b) : Detail Independence of s:

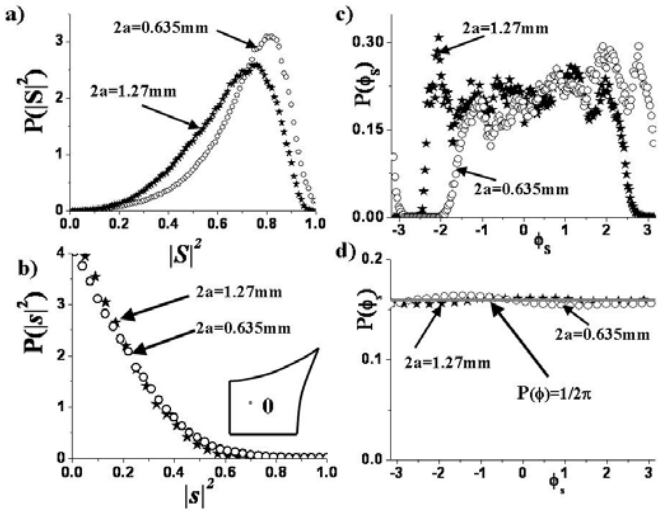

**Fig.3:** (a) PDF for the un-normalized Loss Case 0 cavity $|S|^2$ in the frequency range of 6 to 11.85 GHz for two different coupling antenna diameters 2a = 0.635 mm (circles) and 2a = 1.27 mm (solid stars). (b) PDF for the normalized cavity $|s|^2$ in the frequency range of 6 to 11.85 GHz for two different coupling antenna diameters 2a = 0.635 mm (circles) and 2a = 1.27 mm(solid stars). Note that the disparities seen in Fig. 3.(a) on account of the different coupling geometries disappear after normalization. (c) PDF for the un-normalized cavity phase ($\phi_S$) for Loss Case 0 in the frequency range of 6 to 11.85 GHz for two different coupling antenna diameters 2a = 0.635 mm (circles) and 2a = 1.27 mm (solid stars). (d) PDF for the normalized cavity phase ($\phi_s$) in the frequency range of 6 to 11.85 GHz for two different coupling antenna diameters 2a = 0.635 mm (circles) and 2a = 1.27 mm (solid stars). The normalized phase PDFs for the stars and circles in Fig. 3(d) are nearly uniformly distributed (the gray line in Fig. 4(d) shows a perfectly uniform distribution $P(\phi) = 1/2\pi$ ).

To further verify that the normalized *s* does not include any artifacts of system-specific, non-ideal coupling, we take two identical wave chaotic cavities and change only the inner diameter of the coupling coaxial cable from 2a=1.27 mm to 2a=0.635 mm. Since the modification of the coaxial cable size barely changes the properties of the cavity, we assume that the loss parameters in these two cases are the same. The difference in the coupling geometry manifests itself as gross differences in the distribution of the raw cavity scattering coefficients *S*. This is clearly observable for the PDFs of the cavity power reflection coefficient $|S|^2$ as shown in Fig. 3(a) and the PDFs for the phase of *S* (denoted $\phi_S$) shown in Fig.3(c), for Loss Case 0 over a frequency range of 6 to 11.85 GHz. However, after measurement of the corresponding radiation impedance and the normalization procedure described above, we observe that the PDFs for the normalized power reflection coefficients are nearly identical, as shown in (Fig.3(b)) for $|s|^2$ and the phase ($\phi_s$) in (Fig.3(d)). This supports the theoretical prediction that the normalized scattering coefficient *s* is a universal quantity whose statistics does not depend on the nature of the coupling antenna. Similarly, in Fig.3(c), the phase $\phi_S$ of the cavity scattering coefficient *S* shows preference for certain angles. This is expected because of the non-ideal coupling (impedance mismatch) that exists between the antenna and the transmission line. After normalization, the effects of non-ideal coupling are removed and only the scattering phase of an ensemble of ideally coupled chaotic systems (in which all scattering phases are equally likely) is seen. Hence, consistent with theoretical expectations, the phase $\phi_s$ of normalized *s* data show an approximately uniform distribution (Fig.3(d)).

### III(c) : Variation of s with Loss :

Having established that the coupling geometry is irrelevant for the distribution of *s*, we fix the coupling geometry (coaxial cable with inner diameter 2a=1.27 mm) and vary the degree of quantified loss within the cavity. Three loss cases will be considered, namely, loss case 0, 1 and 3. In Ref.[18] the degree of loss is characterized by a single damping parameter $\tilde{k}^2/Q$. Here, $\tilde{k}^2 = k^2/\Delta k_n^2$, where k=2πf/c is the wave number for the incoming frequency *f* and $\Delta k_n^2$ is the mean spacing of the adjacent eigenvalues $k_n^2$. The quantity $Q$ is the quality factor of the cavity. The parameter $\tilde{k}^2/Q$ represents the ratio of the frequency width of the cavity resonances due to distributed losses, and the average spacing between resonant frequencies.

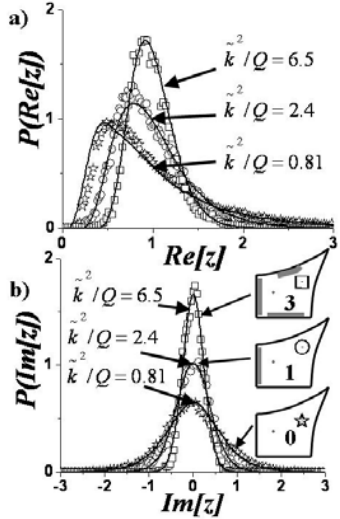

Fig.4: PDFs for the (a) real and (b) imaginary parts of the normalized cavity impedance $z$ for a wave chaotic microwave cavity between 6.5 and 7.8 GHz with $h$ = 7.87 mm and $2a$ = 1.27 mm, for three values of loss in the cavity (open stars: 0, circles: 1, squares: 3 strips of absorber). Also shown are single parameter, simultaneous fits for both PDFs (solid curves), where the loss parameter $\tilde{k}^2/Q$ is obtained from the variance of the data in (a) and (b).

For sufficiently high loss, the variances ($\sigma^2$) of the PDFs of the real and imaginary parts of $z$ can be related to $\tilde{k}^2/Q$ by [18],

$$\sigma^2_{Re(z)} = \sigma^2_{Im(z)} = Q/(\pi \tilde{k}^2). \qquad (7)$$

This relation has been experimentally validated for different cavities and for different coupling geometries [17]. Thus we can determine the damping parameter from measuring the variance of the PDFs of the real or imaginary part of $z$ (such as those shown in Fig. 4). With this parameter determined, we use a Monte-Carlo simulation based on random matrix theory (see Eq.(29) of Ref.[17] and discussion therein) to calculate the theoretically predicted PDFs of $z$ and $s$. (Approximate formulas for the PDFs of $Re[z]$ [21] and $Im[z]$ [26] which agree well with the Monte Carlo results are also available). The solid curves in Fig. 4 are plots from RMT for values of $\tilde{k}^2/Q$ that are obtained by computing the variances, while the symbols are obtained from histogram approximations to the PDFs of the normalized impedance $z$ extracted from the experimental data over a frequency range of 6.5 to 7.8 GHz. Generally, as the loss of the cavity increases, the PDF of the normalized imaginary part of the impedance loses its long tails and begins to sharpen up, developing a Gaussian appearance (Fig.4(b)). At the same time, the PDF of the normalized real part smoothly evolves from being peaked between $Re(z) = 0$ and $Re(z) = 1$, into a Gaussian-like distribution that peaks at $Re(z) = 1$ (Fig.4(a)).

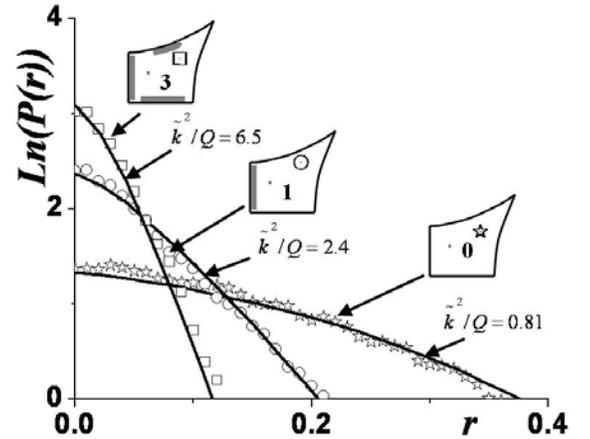

Fig.5: PDF for the normalized power reflection coefficient $r = |s|^2$ on a natural log scale for Loss Case 0, 1, 3 (stars, circles and squares respectively) in the frequency range of 6.5 to 7.8 GHz. These are from the same data as used to obtain Fig.4. Also shown is the prediction of the model in [17] (solid lines) for $P(r)$ using the values of $\tilde{k}^2/Q$ obtained from the variances of the distributions in Fig.4.

The symbols in Fig.5 (presented on a semi-log scale) show the PDF of the normalized power reflection coefficient ($r = |s|^2$) in the frequency range 6.5 to 7.8 GHz for three different loss cases. The solid lines are the predictions for the PDF of $r$, $P(r)$ for different values of the loss parameter $\tilde{k}^2/Q$. In Fig.5, the $\tilde{k}^2/Q$ parameters are the same as those for Fig. 4, and were obtained from the variances of the PDFs of the real or imaginary parts of $z$. We observe that our data conforms well to the predictions from random matrix theory for all degrees of loss.

## IV: Relationship between Cavity and Radiation Power Reflection Coefficients:

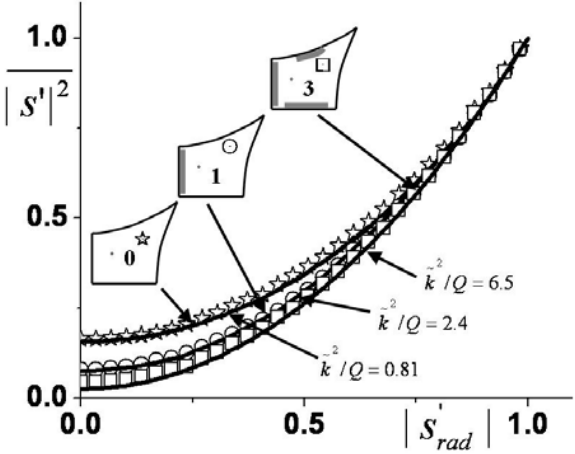

Fig.6: Dependence of the average of the cavity power reflection coefficient $\overline{|s'|^2}$ on the magnitude of the radiation scattering coefficient $|s'_{rad}|$, for different loss cases (Loss Case 0: stars; Loss Case 1: circles; Loss Case 3: squares). The data is shown for the frequency range of 6.5 to 7.8 GHz. Also shown are the numerical simulations from the RMT based upon the $\tilde{k}^2/Q$ values 0.81, 2.4 and 6.5 for loss case 0, 1 and 3 respectively (solid lines).

As a final experimental test, we would like to examine how the measured cavity power reflection coefficient depends only on the radiation scattering coefficient and losses in the cavity. Ref.[18] predicts that the average value of the cavity power reflection coefficient $\overline{|S|^2}$ depends only on the magnitude of the radiation scattering coefficient ($|S_{rad}|$) and the loss parameter $\tilde{k}^2/Q$, and is independent of the phase of $S_{rad}$. The quantity $|S_{rad}|$ is related to the radiation impedance ($Z_{rad} = R_{rad} + iX_{rad}$) through the transformation,

$$|S_{rad}| = \sqrt{\frac{(R_{rad} - Z_o)^2 + (X_{rad})^2}{(R_{rad} + Z_o)^2 + (X_{rad})^2}}. \quad (8)$$

We consider a cavity having quantified loss (Loss Case 0, 1 and 3), with a coupling port of diameter 2a=1.27 mm and over the frequency range of 6.5 to 7.8 GHz. Having experimentally generated the normalized $z$ as described above, we then simulate an ensemble of similar cavities but with different coupling "geometries". This is done by means of a lossless two-port impedance transformation [18] of our $z$ data, as described by the relation,

$$z' = \frac{1}{(1/z + i\beta)}. \quad (9)$$

which corresponds to adding a reactive impedance $-i/\beta$ in parallel with the impedance $z$.

The quantity $z'$ thus simulates the impedance of a hypothetical cavity that is non-ideally coupled to the excitation port, and the coupling geometry is characterized by the real factor $\beta$, which can be varied in a controlled manner. We also define a transformed radiation impedance ($z'_{rad}$) given by,

$$z'_{rad} = \frac{1}{(1 + i\beta)}. \quad (10)$$

For the generation of $z'_{rad}$, the factor $\beta$ is varied over the same range of values as used to generate $z'$. Having determined $z'$ and its corresponding $z'_{rad}$, we determine the scattering coefficients $s'$ and $s'_{rad}$ through the transformations,

$$s' = (z'-1)/(z'+1) \quad (11)$$
$$s'_{rad} = (z'_{rad} - 1)/(z'_{rad} + 1) \quad (12)$$

A range of $\beta$ values are chosen to cover all possible coupling scenarios. We then plot the average of $|s'|^2$ (i.e., $\overline{|s'|^2}$) as a function of $s'_{rad}$. This approach is followed for all three loss cases (Loss case 0, 1 and 3) resulting in the data sets with star, circles and squares, respectively, in Fig.6. First note that all curves originate from the point $|s'|^2 = |s'_{rad}| = 1$, which may be thought of as the perfectly mismatched case. Next consider $|s'_{rad}| < 1$, and observe that as the losses increase, the curves shift downwards for a fixed coupling (characterized by $|s'_{rad}|$). This is intuitively reasonable because, as the absorption (losses) within the cavity increases, we expect less signal to return to the antenna (i.e. smaller $\overline{|s'|}$) for a given coupling $|s'_{rad}|$. From the variance of the PDF of Re[$z$] for the above loss cases, we determine $\tilde{k}^2/Q$ to be 0.81, 2.4 and 6.5 for Loss Case 0,1 and 3 respectively. The solid lines in Fig.6 are obtained from the RMT theoretical predictions for the perfectly coupled scattering coefficient $s$ with the appropriate values for $\tilde{k}^2/Q$. Next, these are transformed using Eqs. (9) and (10) with the same range of coupling factors ($\beta$) as used for the experimentally determined $z'$. We observe good agreement between the numerical simulations from RMT (solid lines in Fig. 6) and our experimentally derived points.

For a given lossy cavity one can also consider its lossless $N$-port equivalent. By the lossless $N$-port equivalent we mean that the effect of the losses distributed in the walls of our cavity can be approximated by a lossless cavity with $N-1$ extra perfectly-coupled (pc) ports through which power coupled into the cavity can leave. The point $|S_{rad}|=|s'_{rad}|=0$ in Fig.6 corresponds to perfect coupling. In this case, Ref. [18] predicts that the vertical axis intercept of these curves corresponds to the lossless $N$-port equivalent of the distributed losses within the cavity; i.e., at $|s'_{rad}|=0$ we have $\overline{|s'|^2}\big|_{pc} = 2/(N+1)$ ( for time-reversal symmetric wave chaotic systems). Thus, in our experiment the quantified loss in Loss Case 0, 1 and 3 is equivalent to ~ 11, 24 and 45 perfectly-coupled ports, respectively. In other words, for all intents and purposes, the cavity can be considered lossless but perfectly coupled to this number of ports.

### V : Validating the Use of Radiation Impedance to Characterize Non-ideal Coupling.

In section III we used the radiation impedance ($Z_{rad}$), or the radiation scattering coefficient ($S_{rad}$), as a tool to characterize the non-ideal coupling (direct processes) between the antenna and the cavity. This quantity is measurable and is only dependent on the local geometry around the port. As previously noted, Refs.[15,16] use configuration and frequency averaged scattering data to obtain an approximation to $<S>$. For a given center frequency $f_0$, this procedure relies on the satisfaction of two requirements: first the range of $\Delta f$ must be large enough to include a large number of modes; second, $S_{rad}$ must vary little over the range of $\Delta f$.

The nature of the variation of $S$ with frequency is illustrated in Fig. 7, where a plot (Re($S$), Im($S$)) of the complex scattering coefficient for a cavity in the frequency range of 6 to 12 GHz is shown. The blue trace shows results for $S$ for a single configuration of the cavity corresponding to a given position and orientation of the perturbers (Fig.1(a)). Isolated resonances are seen as circular loops in the polar plot. The degree of coupling is indicated by the diameter of the loops. Frequency ranges where the coupling is good would manifest themselves as large loops, while those frequency ranges with poor coupling result in smaller loops. Ensemble averaging one hundred such different configurations of this cavity for different positions and orientations of the perturber produces the black trace denoted as $<S>_{100}$. Note that even with one hundred cavity renditions, the fluctuations in $<S>_{100}$ are still present and are seen as the meanders in the black trace. The red trace, which corresponds to the radiation scattering coefficient for this antenna geometry, is devoid of such fluctuations (because there are no reflected waves from the far walls back to the port) and is easily obtainable in practice without resorting to generating large ensemble sets of cavity configurations. Moreover, since the radiation impedance of the port is also a function of frequency, there is no constraint on the frequency span where the analysis for obtaining the universal statistics of $s$ (or $z$) can be carried out.

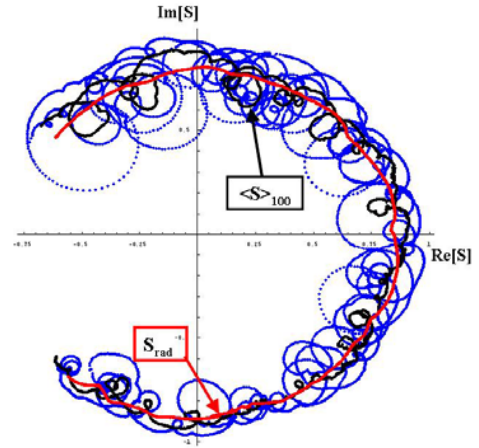

**Fig.7: (color online:)** Polar plot for the cavity scattering coefficient $S = \text{Re}(S) + i\,\text{Im}(S)$ is shown for a frequency range of 6 to 12 GHz for Loss Case 0 and with a coupling port of diameter 2a = 1.27 mm. The blue trace represents one single rendition of the cavity for a selected position and orientation of the perturbers. Each circular loop represents an isolated resonance. The black trace is the ensemble average $<S>_{100}$ over one hundred different locations and orientations of the perturbers within the cavity. The meandering nature of the black trace shows that remnants of the cavity resonances are still present because of the finite number of ensemble averages. The red trace shows the radiation scattering coefficient for the same port. This trace is smooth because the radiation scattering coefficient approximates the cavity boundaries being extended to infinity.

To quantitatively illustrate this point, we simulate the non-universal scattering statistics of a given cavity for a given type of coupling using only the measured radiation impedance of the coupling port and the numerically generated normalized impedance $z$ from RMT, which depends only upon the net losses within the cavity. We consider a Loss Case 0 cavity, over a frequency range of 6 to 7.5GHz, which is excited by means of a coaxial cable of inner diameter (2a=1.27mm). The variation in $|<S>_{100}|$ (inset of Fig.8(b)) indicates that the coupling characteristics for this setup fluctuate over the given frequency range, undergoing roughly four or five oscillations over a range in $|<S>_{100}|$ of order 0.2. Thus the frequency averaged $|<S>_{100}|$ would be expected to be an unreliable estimate to parameterize the coupling over this frequency range.

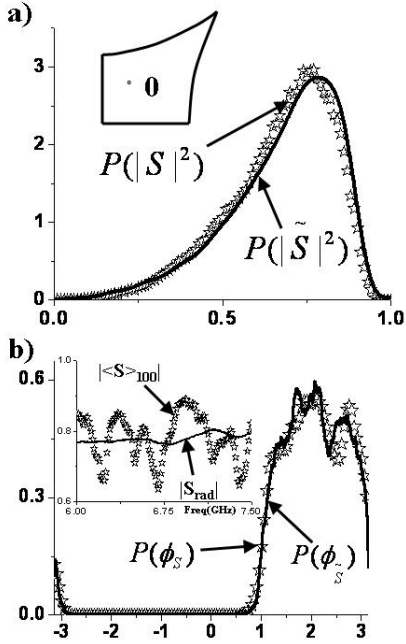

**Fig.8:** (a) The experimental PDF for the Loss Case 0 cavity power reflection coefficient ($|S|^2$) (stars) over a frequency range of 6 to 7.5GHz. Also shown is the numerical estimate $P(|\tilde{S}|^2)$ (solid trace) determined from RMT and the experimentally measured radiation impedance of the port ($Z_{rad}$). (b) The experimental PDF for the Loss Case 0 cavity scattering phase ($\phi_S$) (stars) over a frequency range of 6 to 7.5GHz. Also shown is the numerical estimate $P_{(\phi_{\tilde{S}})}$ (solid trace) determined from RMT and the experimentally measured radiation impedance of the port. We observe good agreement between the measured data and the numerically estimated PDFs. The inset shows the fluctuation in $|<S>_{100}|$ (stars) over the frequency range of 6 to 7.5GHz, while the solid trace shows the magnitude of the experimentally measured radiation scattering coefficient ($|S_{rad}|$).

We can estimate the parameter $\tilde{k}^2/Q$ using the center frequency ($k = 141.3 m^{-1}$), the average spacing between eigenmodes for our cavity ($\Delta k_n^2 = 109.2 m^{-2}$) [22], and typical $Q$ values of ($Q \sim 225$), yielding an estimated $\tilde{k}^2/Q = 0.8$. We use this parameter to generate an ensemble of $s(\omega)$ following Ref.[18] and Eq.(4), combine it with the measured $S_{rad}(\omega)$ of the antenna, and employ Eq.(5) to obtain an estimate of the non-universal system-specific scattering coefficient, which we denote as $\tilde{S}$.

In Fig.8(a), the PDF of $|\tilde{S}|^2$ is shown as the solid trace, while the experimentally measured PDF of $|S|^2$ is shown as the stars. While in Fig. 8(b), the PDF of $\phi_{\tilde{S}}$ is shown as the solid trace with experimentally measured PDF of $\phi_S$ shown as the stars. We observe good agreement between the numerically generated estimate and the actual data. This result validates the use of the radiation impedance (scattering coefficient) to accurately parameterize the system-specific, non-ideal coupling of the ports and also provides us with a way to predict beforehand the statistical properties of other complicated enclosures non-ideally coupled to external ports.

### *VI : Conclusions :*

The results discussed in this paper serve to establish a solid ground that can be used to extract universal statistical properties from data on wave chaotic systems, or to engineer wave chaotic cavities with specific statistical transport properties. In addition, given the frequency, volume and amount of losses (parameterized by $Q$) within the enclosure, and the radiation impedance of the ports, Refs. [17] and [18] provide us with a tool to predict the statistics of the cavity response ($Z$ and $S$) *a priori*.

We have shown that a simple normalization based on the radiation impedance can be used to remove non-universal, system-specific coupling details and bring out the universality in the measured impedance and scattering statistics of wave chaotic systems. This normalization procedure has allowed us to experimentally verify theoretical predictions for the universal properties of a one-port wave chaotic system. We have also tested several aspects of the theory in the realm of intermediate to high loss and for different coupling geometries and find good agreement with theoretical predictions. We have shown that the average of the cavity power reflection coefficient $\overline{|S|^2}$ depends only on the magnitude of the radiation scattering coefficient $|S_{rad}|$ and the degree of loss, and have obtained good agreement between theory and experiments. Finally, we also demonstrate the ability of this normalization procedure to faithfully reproduce the non-universal statistics of the scattering coefficient phase and magnitude of chaotic cavities when $|<S>|$ is not constant over the frequency range examined. These results should not be regarded as limited to microwave cavities or any specific coupling structure, but as applying to any wave chaotic system coupled to the outside world.

**Acknowledgements:**

We acknowledge useful discussions with R. Prange and S. Fishman, as well as comments from Y. Fyodorov, D.V. Savin and P. Brouwer. This work was supported by the DOD MURI for the study of microwave effects under AFOSR Grant F496200110374 and an AFOSR DURIP Grant FA95500410295.